\documentclass[12pt,aps,prd,preprint,tightenlines,superscriptaddress,showpacs,nofootinbib]
{revtex4-1}
\pdfoutput=1
\usepackage{amsmath,amssymb,amsthm,amsfonts}
\usepackage{slashed} 
\usepackage{graphicx}
\usepackage{caption}
\usepackage{subcaption}
\usepackage{dcolumn}
\usepackage{hyperref}      
\usepackage{bm}
\usepackage{epsfig}
\usepackage{epstopdf}
\usepackage{setspace}
\usepackage{color}
\usepackage{xcolor,}
\usepackage{multirow}
\newcommand{\beq}{\begin{equation}}
\newcommand{\eeq}{\end{equation}}
\newcommand{\bea}{\begin{eqnarray}}
\newcommand{\eea}{\end{eqnarray}}
\newcommand{\beqa}{\begin{eqnarray}}
\newcommand{\eeqa}{\end{eqnarray}}

\begin{document}

\title{The $\tau$ Magnetic Dipole Moment at Future Lepton Colliders\vspace*{0.25cm}}

\author{Jessica N. Howard}
\email{jnhoward@uci.edu}
\affiliation{Department of Physics and Astronomy, University of
California, Irvine, CA 92697, USA \vspace*{1cm}}

\author{Arvind Rajaraman}
\email{arajaram@uci.edu }
\affiliation{Department of Physics and Astronomy, University of
California, Irvine, CA 92697, USA \vspace*{1cm}}

\author{Rebecca Riley}
\email{rriley1@uci.edu}
\affiliation{Department of Physics and Astronomy, University of
California, Irvine, CA 92697, USA \vspace*{1cm}}

\author{Tim M. P. Tait}
\email{ttait@uci.edu}
\affiliation{Department of Physics and Astronomy, University of
California, Irvine, CA 92697, USA \vspace*{1cm}}

\preprint{UCI-HEP-TR-2018-11}

\begin{abstract}
The magnetic moment of the $\tau$ lepton is an interesting quantity that is potentially sensitive to
physics beyond the Standard Model.  Electroweak gauge invariance implies that a heavy new physics
contribution to it takes the form of an operator which involves the Higgs boson, implying that rare
Higgs decays are able to probe the same physics as $a_\tau$.  We examine the prospects for
rare Higgs decays at future high energy lepton (electron or muon) colliders, and find that
such a project collecting a few ab$^{-1}$
would be able to advance our understanding of this physics by roughly a factor of 10
compared to the expected reach of the high luminosity LHC.
\end{abstract}

\maketitle

\section{Introduction}

Anomalous magnetic moments of charged fermions occupy a special role in our understanding of the Standard Model (SM) of 
particle physics.  Early on, the tiny electromagnetic correction to the electron's magnetic moment provided one of the first
indications that quantum field theory was the correct language to describe subatomic physics.  More recently, the anomalous
magnetic moment of the muon is one of the few experimental measurements to stubbornly resist being well-described
by the Standard Model\footnote{It also bears mentioning that puzzling results for the
proton radius extracted from muonic hydrogen \cite{Pohl:2010zza} and indications for
lepton-non-universality in semi-leptonic $B$ decays 
\cite{Aaij:2014pli,Aaij:2015esa,Aaij:2015oid,Aaij:2016flj,Wehle:2016yoi} may further point to new physics influencing
muonic observables.} \cite{Bennett:2006fi}.  A short distance contribution to the anomalous magnetic moment
takes the form of a dimension-5 operator,
\beq
a_\psi~ \frac{e }{2 m_\psi}~ \bar \psi  \sigma^{\mu\nu}\psi ~ F_{\mu\nu}
\label{eq:general_dipole_lowE}
\eeq
which implies a chiral-flip on the fermion.  Since chirality change involves an insertion of the mass, it is natural to speculate that
if there is new physics subtly influencing the muon magnetic moment, 
it should manifest even more strongly for the tau lepton.

The electroweak $SU(2) \times U(1)$ gauge symmetry would forbid chirality-changing interactions of the SM fermions.  Consequently,
their presence ultimately derives from the fact that the symmetry is spontaneously broken by the vacuum expectation value (VEV)
of the Higgs.  Indeed, promoting the tau magnetic moment operator into an $SU(2) \times U(1)$ invariant form leads to a pair
of dimension six terms,
\beq
c_1 ~\bar \tau_R \sigma^{\mu\nu}B_{\mu\nu} ~H^\dagger L_3
+ c_2 ~\bar \tau_R \sigma^{\mu\nu}~H^\dagger W_{\mu\nu}  L_3  
+ h.c.
\label{eq:dim6}
\eeq
where $L_3$ is the left-handed $SU(2)$ lepton doublet containing $\tau_L$, 
$H$ is the Higgs doublet, $B_{\mu \nu}$ and $W_{\mu \nu}$ are the field strengths for
the hypercharge and $SU(2)$ gauge bosons,
and $c_1$ and $c_2$ are (generically complex) coefficients with units of $(\rm{energy})^{-2}$ which encapsulate
the residual effects of heavy physics at low energies.  
Replacing the Higgs
with its VEV generates the magnetic dipole moment (and also generically modify
$Z$ boson couplings and, if $c_1$ and $c_2$ are complex, contributes to the electric dipole).  But such interactions {\em necessarily}
imply a modification of the coupling of the Higgs to $\tau^+ \tau^- \gamma$ as well,
\beq
{1\over \Lambda^2} h~ \bar \tau \sigma^{\mu\nu}~\tau F_{\mu \nu},
\label{eq:mdm}
\eeq
where $h$ is the field corresponding to the Higgs boson and $\Lambda$ is a linear combination of the original couplings $c_1$ and $c_2$.
Thus, rare Higgs decays offer a complementary, high energy probe of anomalous magnetic moments.  Guided by the expectation that
new physics may be more evident for the tau, we focus on the operator described in Equation~(\ref{eq:mdm}).

Our current understanding of the tau magnetic dipole moment is relatively modest.
Previous experiments have focused on precision measurements of the $\tau$ itself~\cite{
Silverman:1982ft,
Almeida:1991hq,
delAguila:1991rm,
Samuel:1992fm,
Aeppli:1992tp,
Escribano:1993pq, 
Escribano:1996wp,
GonzalezSprinberg:2000mk,
Bernabeu:2007rr,
Bernabeu:2008ii,
Atag:2010ja,
Peressutti:2012zz,
Hayreter:2013vna,
Fael:2013ij,
Hayreter:2015cia,
Eidelman:2016aih},
with the strongest constraints on these operators coming from the kinematics of the
production process $e^+ e^- \rightarrow \tau^+ \tau^-$ at LEP2~\cite{Abdallah:2003xd,Heister:2002ik}, which places
the limit:
\bea
-0.052 < a^\gamma_\tau < 0.013,\quad 95\%~{\rm CL},
\label{eq:agamma_bound}
\eea
corresponding to \cite{Galon:2016ngp}:
\beq
|\Lambda| > \begin{array}{l} 333~ \rm{GeV} \\
666~\rm{GeV}  \end{array}
~~~~~~~~~{\rm for:}~~~
\begin{array}{r}  ~\Lambda^2 > 0\\
 \Lambda^2 < 0 \end{array}.
\label{eq:LEP_bounds}
\eeq

At higher energies, future lepton colliders can directly produce the Higgs boson, and look for the rare Higgs decay
\beq
h \to \tau^+\tau^- \gamma ,
\label{eq:3body_decay}
\eeq
in which the new physics contribution interferes with the SM decay $h \rightarrow \tau^+ \tau^-$ followed by radiation
of a hard photon from one of the $\tau$'s.
Constraints on the size of this rare decay imply a bounds on the size of $\Lambda$, and hence on the magnetic moment.
This process was previously studied in Ref.~\cite{Galon:2016ngp}, and found to be promising for the end stage LHC
running in a high luminosity mode.  In this work, we extend this study to future lepton colliders and examine the
contribution such machines could contribute to our understanding of the tau magnetic moment.

\section{Future Lepton Colliders}

\begin{figure}[t!]
\includegraphics[width=.85\textwidth]{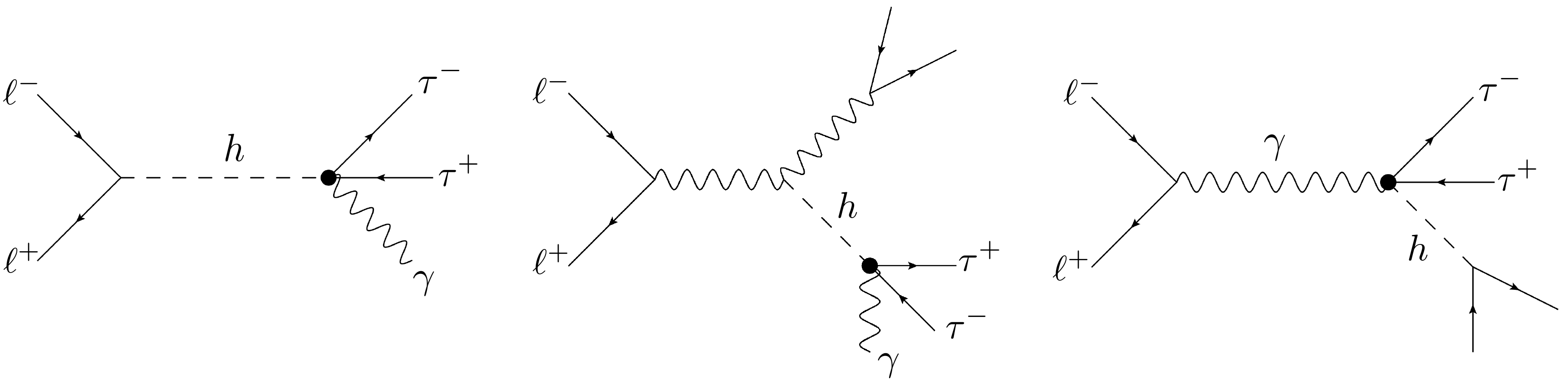}
\caption{Representative Feynman diagrams for the processes
$\mu^+ \mu^- \rightarrow h^{(*)} \rightarrow \tau^+ \tau^- \gamma$,
$\ell^+ \ell^- \rightarrow Z^* \rightarrow Z h$ followed by $h \rightarrow \tau^+ \tau^- \gamma$, and
$\ell^+ \ell^- \rightarrow \gamma^* \rightarrow \tau^+ \tau^- h$ (left to right).}
\label{fig:feynman}
\end{figure}

The operator of Equation~(\ref{eq:mdm}) leads to exotic processes involving a Higgs boson, a photon, and a $\tau^+ \tau^-$ pair.  At a future
high energy lepton collider, there are a number of potential ways to search for its presence.  We consider three reactions which can
probe this process:
\begin{itemize}
\item $\mu^+ \mu^- \rightarrow h^{(*)} \rightarrow \tau^+ \tau^- \gamma$;
\item $\ell^+ \ell^- \rightarrow Z^* \rightarrow Z h$ followed by $h \rightarrow \tau^+ \tau^- \gamma$; or
\item $\ell^+ \ell^- \rightarrow \gamma^* \rightarrow \tau^+ \tau^- h$ followed by any SM Higgs decay.
\end{itemize}
Representative Feynman diagrams for the three processes are shown in Figure~\ref{fig:feynman}.  The first two processes probe the
$\tau$ magnetic dipole moment via Higgs decay, and thus typically have momentum transfer characterized by $m_h$.  The third process
involves a momentum transfer of order $\sqrt{s}$, which may be important at very high energies.  In these reactions $\ell \equiv e, \mu$
are relevant both at future electron and future muon colliders, whereas the first process is very suppressed at an $e^+ e^-$ machine
due to the tiny electron Yukawa interaction.

\subsection{$\ell^+ \ell^- \rightarrow Z^* \rightarrow Z h$}

We find that for intermediate collider energy, the process $\ell^+ \ell^- \rightarrow Z^* \rightarrow Z h$ provides the most stringent
constraint, with $\ell^+ \ell^- \rightarrow \gamma^* \rightarrow \tau^+ \tau^- h$ becoming comparable at very high collider
energies. For the energies considered, the process $\mu^+ \mu^- \rightarrow h^{(*)} \rightarrow \tau^+ \tau^- \gamma$
always involves an off-shell Higgs boson, and is never the dominant probe.
We therefore focus our discussion on the $Z h$ production mode, followed by $h \rightarrow \tau^+ \tau^- \gamma$, leading to a
$Z \tau^+\tau^-\gamma$ final state.  

The detailed experimental reconstruction depends rather sensitively on the design and performance of the
detectors, and are not currently completely well-defined.  
However, all current plans propose very precise detectors providing exquisite information 
via calorimetric and tracking metrics, which are likely to reconstruct all visible particles produced in a collision.
For this reason, we we eschew a specific detector simulation and instead work at the parton level, imposing stiff
reconstruction cuts such that we expect the $\tau$'s, $\gamma$'s, and $Z$ bosons (for visible decays) can be reconstructed
with near-perfect precision,
\bea
p_T^{\gamma,\tau} \geq 10~{\rm GeV}, ~~~~~~|\eta^{\gamma,\tau}| \leq 2.5,
\eea
where $p_T$ and $\eta$ are the transverse momentum and pseudo-rapidity, respectively.
Specifically, these stiff cuts on the $\tau$s insure that their decay products are collimated which allows their momenta to be
reconstructed reasonably accurately, despite the unmeasured energy going into neutrinos
(for a more detailed discussion of the requirements in a similar context, 
see Ref. \cite{Murakami:2014tna}).
We simulate both signal and SM background processes using
MadGraph5\_aMC@NLO (MG5)~\cite{MG5}, with the
FeynRules SM implementation~\cite{Degrande:2011ua,Alloul:2013bka,Alloul:2013naa},
supplemented with the tau dipole operators of Eq.~(\ref{eq:dim6}).

\subsubsection{Background and Selection Cuts}

There are multiple SM background processes contributing to the $Z \tau^+\tau^-\gamma$ signature.  These are primarily:
\begin{itemize}
\item radiation of a photon from the initial state of $\ell^+ \ell^- \rightarrow ZZ$ followed by one of the
$Z$ bosons decaying into $\tau^+ \tau^-$;
\item initial state photon radiation from $\ell^+ \ell^- \rightarrow Z h$, where $h \rightarrow \tau^+ \tau^-$; and
\item final state radiation of a photon from a $\tau$ in the process $\ell^+ \ell^- \rightarrow Z Z$ where one $Z$ decays to
$\tau^+ \tau^- \gamma$.
\end{itemize}
The initial state radiation contributions to the background are slightly different for electron and muon beams, because of
the dependence on the lepton mass in the collinear region of kinematics.

The primary tool to sift the new physics signal from these backgrounds is to reconstruct the invariant masses $M_{\tau \tau}$
and $M_{\tau \tau \gamma}$.  In order to remove backgrounds where the $\tau$s are produced by off-shell photons
or close to on-shell $Z$ bosons, we exclude events for which,
\bea
M_{\tau\tau} \leq 10~{\rm GeV} ~~~~~~{\rm and}~~~~~~ 80~{\rm GeV} \leq M_{\tau\tau} \leq 100~{\rm GeV}.
\eea
We would also like to avoid processes for which the $\tau$s are produced by an on-shell Higgs boson decay, with the
additional photon radiated from the initial state.  This is accomplished by vetoing events which satisfy,
\bea
120~{\rm GeV} \leq M_{\tau\tau} \leq 130~{\rm GeV}.
\eea
We further select events for which the $\tau \tau \gamma$ are consistent with coming from an on-shell Higgs decay,
requiring,
\bea
120~{\rm GeV} \leq M_{\tau\tau\gamma} \leq 130~{\rm GeV}.
\eea
Our choice of cuts is conservative in the sense that they assume windows around the $Z$ and Higgs masses which are
${\cal O}(10\%)$, far larger than the expected order per cent level
experimental resolution for a realistic detector or the intrinsic widths of
the $Z$ and Higgs bosons. 
In Table~\ref{table:cut_list_baseline}, we show the cross sections before and after cuts for simulations containing the
SM alone, and for the SM plus the new physics operator for two choices of
\bea
\alpha \equiv \frac{G_F}{\Lambda^2},
\eea
$\alpha = 0.1$, and $\alpha=0.2$ for the representative case of an $e^+ e^-$ collider operating at 
a center-of-mass energy $\sqrt{s} = 500$~GeV.

\begin{table}[t!]
\begin{tabular}{c|c|c|c}
 & 
~~~~~~~~~~~~SM~~~~~~~~~~~~~ & 
~~~~~~~~~$\alpha = 0.1$~~~~~~~~~& 
~~~~~~~~~$\alpha = 0.2$~~~~~~~~~
\\\hline
~~~Before cuts~~~
& $1.90 \times 10^{-3}$~pb
& $2.55 \times 10^{-3}$~pb
& $4.43 \times 10^{-3}$~pb \\
$M_{\tau\tau}$ cuts
& $6.66 \times 10^{-4}$~pb 
& $8.98 \times 10^{-4}$~pb
& $1.60 \times 10^{-3}$~pb \\
$M_{\tau\tau\gamma}$ cuts
& $3.34 \times 10^{-5}$~pb
& $1.30 \times 10^{-4}$~pb
& $4.55 \times 10^{-4}$~pb
\\
\end{tabular}
\caption{Cross sections before and after cuts for the process $e^+ e^- \rightarrow Z \tau^+\tau^-\gamma$
at $\sqrt{s} = 500$~GeV in the SM and for two choices of $\alpha$, as indicated.}
\label{table:cut_list_baseline}
\end{table} 

\subsubsection{Analysis}

We extract the sensitivity to the $\tau$ magnetic dipole operator by writing the amplitude for the signal process
with the dependence on $\alpha$ explicitly factored out,
\beq
{\cal M}_{\rm sig} = {\cal M}_{\rm{SM}} + \alpha {\cal M}_{\rm{NP}},
\eeq
for which the cross-section (after cuts) for a specific $\sqrt{s}$ takes the form
\beqa
\sigma (\alpha) &=& 
\sigma_{\rm{SM}} + 
2 \alpha~
{\sigma}_{\rm INT}
+ \alpha^2~
{\sigma}_{\rm NP}.
\label{eq:Nsig}
\eeqa
where $\sigma_{\rm{SM}}$, ${\sigma}_{\rm INT}$, and ${\sigma}_{\rm NP}$ represent the SM cross section, interference term,
new physics contributions, respectively.  They are extracted for
$\sqrt{s} = 500$~GeV from the bottom row of Table~\ref{table:cut_list_baseline}, and likewise for other energies simulated. 

\begin{table}[t!]
\begin{tabular}{c|c|c|c} 
~~~~~~$\sqrt{s}$~~~~~~ & ~~~~~~$\alpha=0.001$~~~~~~
&~~~~~~$\alpha=0.045$~~~~~~ & ~~~~~~$\alpha=0.1$~~~~~~ \\
\hline
 250 GeV
& $7.24  \times 10^{6} $  
& $3.28  \times 10^{3} $
& $6.36  \times 10^{2} $ \\
500 GeV
& $2.66  \times 10^{7} $ 
& $5.51 \times 10^{3}$
& $3.51  \times 10^{2}$ \\
600 GeV
& $1.53 \times 10^{6}$
& $2.66 \times 10^{4}$ 
& $3.08 \times 10^{2}$    \\
800 GeV
& $7.35 \times 10^{5}$  
& $4.57 \times 10^{3}$ 
& $2.31 \times 10^{2}$ \\
1000 GeV
& $3.71 \times 10^{5}$ 
& $3.04 \times 10^{3}$ 
& $2.18 \times 10^{2}$ 
\end{tabular}
\caption{Integrated luminosity (in fb$^{-1}$) required for a $5\sigma$ deviation for different center of mass
energies and values of $\alpha$ at an $e^+ e^-$ collider.
\label{tab:epem}}
\end{table}

\begin{table}[t!]
\begin{tabular}{c|c|c|c} 
~~~~~~$\sqrt{s}$~~~~~~ & ~~~~~~$\alpha=0.001$~~~~~~
&~~~~~~$\alpha=0.045$~~~~~~ & ~~~~~~$\alpha=0.1$~~~~~~ \\
\hline
250 GeV
& $8.63 \times 10^{6}$
& $3.98 \times 10^{3}$
& $7.75 \times 10^{2}$\\
500 GeV
& $2.92 \times 10^{6} $
& $2.60 \times 10^{5} $
& $4.59 \times 10^{2} $\\
600 GeV
& $1.23 \times 10^{6} $
& $2.40 \times 10^{5} $
& $3.45 \times 10^{2} $\\
800 GeV
& $4.22 \times 10^{5} $
& $2.94 \times 10^{4} $
& $2.48 \times 10^{2} $\\
1000 GeV
& $8.04 \times 10^{4} $
& $1.56 \times 10^{4} $
& $2.46 \times 10^{2} $\\
\end{tabular}
\caption{Integrated luminosity (in fb$^{-1}$) required for a $5\sigma$ deviation for different center of mass
energies and values of $\alpha$ at a $\mu^+ \mu^-$ collider.
\label{tab:mupmum}}
\end{table}

\begin{figure}[t!]
\includegraphics[width=.95\textwidth]{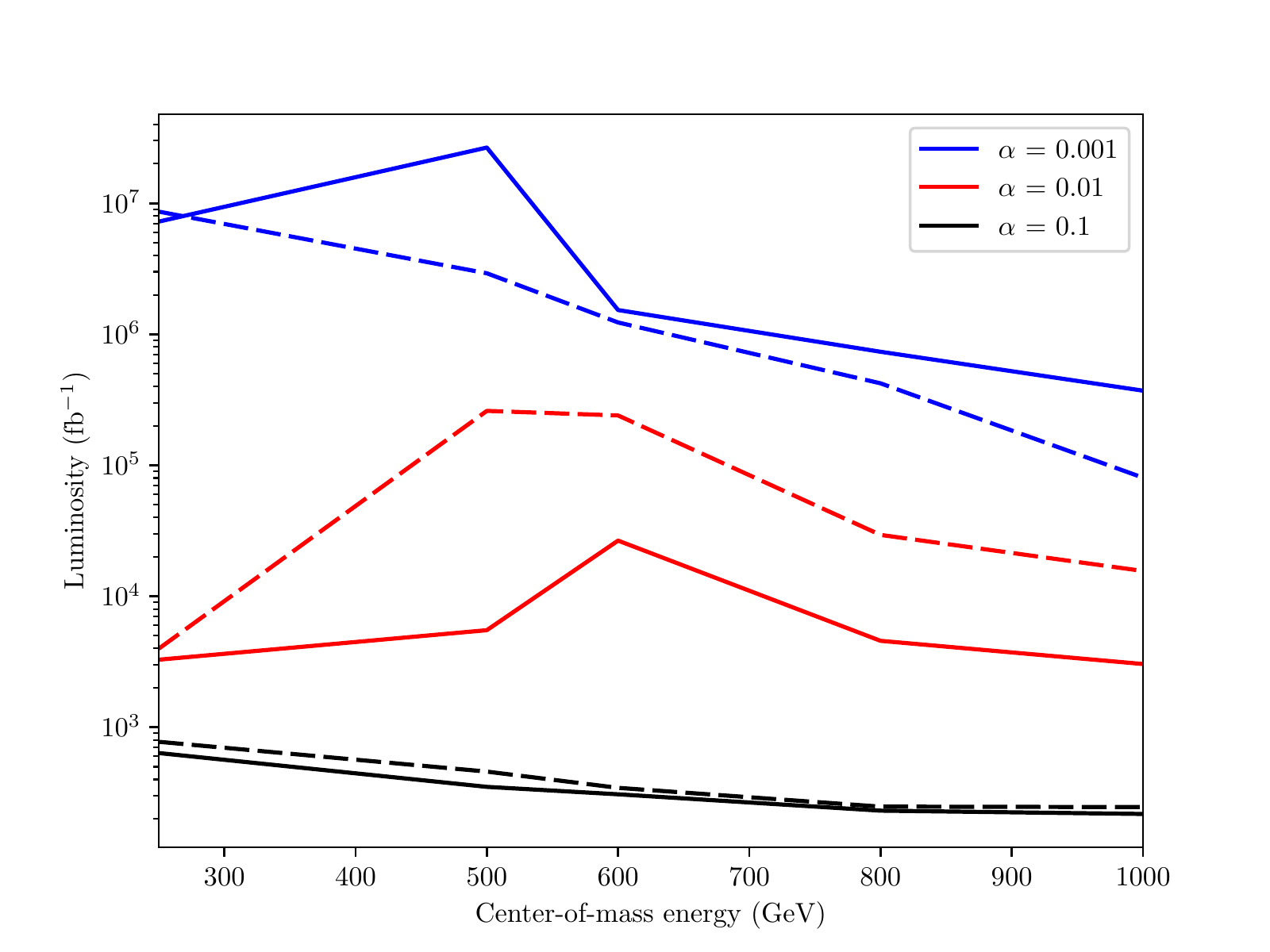}
\caption{Integrated luminosity required for a $5\sigma$ discovery as a function of center of mass energy at a future
$e^+ e^-$ (solid lines) or $\mu^+ \mu^-$ (dashed lines) collider for the indicated values of $\alpha$.}
\label{fig:epem}
\end{figure}

From here, we determine the required luminosity such that the difference between the 
number of events predicted for the Standard Model $B$ and the total number of expected events
for a given $\alpha$, $S+B$ corresponds to a $5\sigma$ statistically significant
deviation,  ${S\over\sqrt{S+B}}>5$ in the regime of Gaussian statistics, $S+B>5$.
Tables~\ref{tab:epem} and \ref{tab:mupmum} show the required luminosity 
in fb$^{-1}$ for a $5\sigma$ deviation at an $e^+e^-$ and $\mu^+\mu^-$ collider, for various
energies and values of $\alpha$.  
The same results are presented graphically in Figure~\ref{fig:epem}.
Evident from the tables, values of $\Lambda$ of order a few
TeV can be probed with datasets of order hundreds of fb$^{-1}$ for $\sqrt{s} \gtrsim 500$~GeV, 
whereas reaching $\Lambda$ of order 10s of TeV requires a few ab$^{-1}$ and $\sqrt{s} \gtrsim 800$~GeV.

\subsection{$\ell^+ \ell^- \rightarrow \gamma^* \rightarrow \tau^+ \tau^- h$}

At higher collider energies, the process $\ell^+ \ell^- \rightarrow \gamma^* \rightarrow \tau^+ \tau^- h$ can take advantage
of the larger $s$-channel momentum transfer, resulting in a larger lever arm when studying short distance physics.
We simulate this process, assuming that the Higgs can be reconstructed with near-perfect efficiency regardless of its
decay mode, and apply selection cuts to the $p_T$ and $\eta$ of the tau leptons and photons
as before.  We select high momentum transfer by retaining events for which
\bea
M_{\tau \tau} \geq 100~{\rm GeV}.
\eea
Using the same criteria as before, we determine the required integrated luminosity for a $5\sigma$ deviation
for a variety of center of mass energies and coupling strengths $\alpha$.
In Tables~\ref{tab:proc2a} and \ref{tab:proc2b}, we show the results for $e^+ e^-$ and $\mu^+ \mu^-$ colliders,
respectively.  These results indicate that an ${\cal O}(10~{\rm TeV})$ lepton collider would provide a {\em very} effective
probe of new physics relevant for the tau magnetic dipole moment.

\begin{table}[t!]
\begin{tabular}{c|c|c|c} 
~~~~~~$\sqrt{s}$~~~~~~ & ~~~~~~$\alpha=0.001$~~~~~~
&~~~~~~$\alpha=0.045$~~~~~~ & ~~~~~~$\alpha=0.1$~~~~~~ \\
\hline
1 TeV
& $1.05  \times 10^{5} $  
& $5.37  \times 10^{1} $
& $1.09  \times 10^{1} $ \\
5 TeV
& $1.90  \times 10^{3} $ 
& $9.36 \times 10^{-1}$
& $1.90  \times 10^{-1}$ \\
10 TeV
& $4.72 \times 10^{2}$
& $2.33 \times 10^{-1}$ 
& $4.72 \times 10^{-2}$  
\end{tabular}
\caption{Integrated luminosity (in fb$^{-1}$) required for a $5\sigma$ deviation for different center of mass
energies and values of $\alpha$ at an $e^+ e^-$ collider.
\label{tab:proc2a}}
\end{table}

\begin{table}[t!]
\begin{tabular}{c|c|c|c} 
~~~~~~$\sqrt{s}$~~~~~~ & ~~~~~~$\alpha=0.001$~~~~~~
&~~~~~~$\alpha=0.045$~~~~~~ & ~~~~~~$\alpha=0.1$~~~~~~ \\
\hline
1 TeV
& $2.19  \times 10^{5} $  
& $2.44  \times 10^{2} $
& $5.08  \times 10^{1} $ \\
5 TeV
& $1.91 \times 10^{3} $ 
& $9.42 \times 10^{-1}$
& $1.91  \times 10^{-1}$ \\
10 TeV
& $4.69 \times 10^{2}$
& $2.32 \times 10^{-1}$ 
& $4.69 \times 10^{-2}$  
\end{tabular}
\caption{Integrated luminosity (in fb$^{-1}$) required for a $5\sigma$ deviation for different center of mass
energies and values of $\alpha$ at an $\mu^+ \mu^-$ collider.
\label{tab:proc2b}}
\end{table}

\section{Conclusions and Outlook}
\label{sec:conclusions}

Given the longstanding discrepancy between the measurement
of the anomalous magnetic moment of the muon 
and its SM predictions, it is natural to wonder if $a_\tau$ might show a related discrepancy
enhanced by its larger mass.  The electroweak-violating nature of the operators contributing to $a_\tau$
imply that an indirect way to access it is through rare Higgs decays into $\tau^+ \tau^- \gamma$.

We have examined the prospects for such a measurement at a future high energy lepton collider,
simulating signal and SM backgrounds (including the important interference between the two) at the
parton level.
We find that at $\sqrt{s} \gtrsim 800$~GeV and having collected a few ab$^{-1}$ of integrated luminosity, it is
likely that such a machine can probe this operator to about a factor of 10 better than the expectations at 
the high luminosity LHC.
Given these initial promising results, it would be worthwhile
to follow up this study with one based on more realistic detector simulations.

\section*{Acknowledgments}

The work of AR and TMPT is supported in part by National Science Foundation 
grant PHY-1620638. Code used for selection cuts and various other aspects of this analysis is located at https://github.com/rebecca-riley/muon\_collider.

\bibliography{htautauaBIB}
\end{document}